\documentclass[nofootinbib,twocolumn,showpacs,preprintnumbers,pre,aps,superscriptaddress]{revtex4-2}

\usepackage[dvipdfmx]{graphicx}
\usepackage{bm}
\usepackage{amsmath}
\usepackage{amssymb}
\usepackage{txfonts}
\usepackage{color}
\usepackage{hyperref}
\usepackage{tikz}
\usepackage{esint}
\usepackage{float} 
\allowdisplaybreaks[1]

\newcommand{\D}{{\rm d}}
\newcommand{\bfx}{{\mathbf x}}

\newcommand{\bfq}{{\mathbf q}}
\newcommand{\bfQ}{{\mathbf Q}}
\newcommand{\bfa}{{\mathbf a}}
\newcommand{\bfA}{{\mathbf A}}

\begin{document}

\title{Covariant Onsager and Onsager-Machlup principles for active and inertial dynamics}

\author{Kento Yasuda}\email{yasuda.kento@nihon-u.ac.jp}
\affiliation{Laboratory of Physics, College of Science and Technology, Nihon University, Funabashi, Chiba 274-8501, Japan}

\author{Bin Zheng}
\affiliation{Zhejiang Key Laboratory of Soft Matter Biomedical Materials, Wenzhou Institute, 
University of Chinese Academy of Sciences, 
Wenzhou, Zhejiang 325001, China}

\author{Zhongqiang Xiong}
\affiliation{Zhejiang Key Laboratory of Soft Matter Biomedical Materials, Wenzhou Institute, 
University of Chinese Academy of Sciences, 
Wenzhou, Zhejiang 325001, China}

\author{Zhanglin Hou}
\affiliation{Zhejiang Key Laboratory of Soft Matter Biomedical Materials, Wenzhou Institute, 
University of Chinese Academy of Sciences, 
Wenzhou, Zhejiang 325001, China}

\author{Kenta Ishimoto}
\affiliation{Department of Mathematics, Kyoto University, Kyoto 606-8502, Japan}

\author{Xinpeng Xu}
\affiliation{Department of Physics and MATEC Key Lab, 
Guangdong Technion - Israel Institute of Technology, Shantou, Guangdong 515063, China}
\affiliation{Technion - Israel Institute of Technology, Haifa 32000, Israel}

\author{David Andelman}\email{andelman@tauex.tau.ac.il}
\affiliation{School of Physics and Astronomy \& Center for Physics and Chemistry of Living Systems, 
Tel Aviv University, Ramat Aviv 69978 Tel Aviv, Israel}

\author{Shigeyuki Komura}\email{komura@wiucas.ac.cn}
\affiliation{Zhejiang Key Laboratory of Soft Matter Biomedical Materials, Wenzhou Institute, 
University of Chinese Academy of Sciences, 
Wenzhou, Zhejiang 325001, China}

\begin{abstract}
The Onsager principle provides a variational route to the phenomenological equations of dissipative 
dynamics through the minimization of the Rayleighian.
We develop a covariant formulation of the Onsager principle for active and inertial systems, 
ensuring geometric consistency under coordinate transformations. 
To further incorporate thermal fluctuations, we formulate the Onsager-Machlup principle for active 
and inertial systems by considering the Onsager-Machlup functional and the corresponding path probability for 
stochastic trajectories.  
Requiring that the path probability obeys the detailed fluctuation theorem, we show that the extended 
Onsager-Machlup theory is consistent with stochastic thermodynamics.
The extended OP and OMP offer a unified and useful variational framework for deriving the dynamical 
equations of active and inertial systems.
\end{abstract}

\maketitle

\section{Introduction}
\label{Intro}

In his seminal 1931 work, Onsager formulated the theory of irreversible thermodynamics, based on phenomenological relations 
in which fluxes are proportional to thermodynamic forces~\cite{Onsager31a,Onsager31b}.
In this framework, he showed that the phenomenological transport coefficient matrix should be symmetric,
a property known as Onsager's reciprocal relations~\cite{DoiBook}.
Onsager's theory was later given a microscopic foundation by Kubo's linear response theory~\cite{Kubo57,KuboYokota57}, 
in which transport coefficients and susceptibilities are expressed in terms of equilibrium time-correlation functions.

Onsager also formulated a variational principle from which the phenomenological equations can be derived in terms of energy 
dissipation~\cite{Onsager31a,Onsager31b}.
By introducing a quantity called the Rayleighian, Doi further developed this variational approach 
and referred to it as the Onsager principle (OP)~\cite{DoiBook}.
For isothermal systems, the Rayleighian typically consists of a dissipation function and the 
rate of change of the free energy.
The OP has been applied to various soft matter systems, such as colloidal suspensions~\cite{Doi11,DoiBook}, 
polymers~\cite{Zhou18,Doi21}, binary mixtures~\cite{Xu15}, liquid crystals~\cite{Doi11}, bilayer membranes and
vesicles~\cite{Fournier15,Okamoto16,Oya18}, and spreading of liquid droplets~\cite{Man17,Hu17}.  
More recently, the OP has been extended to active systems~\cite{Zhang20,Wang21,Ackermann23}, 
in which energy is consumed for self-propulsion or other non-equilibrium dynamics.

In 1953, Onsager and Machlup generalized the OP to incorporate thermal fluctuations in a path-integral 
formulation~\cite{Onsager53,Machlup53b}. 
In what is now called the Onsager-Machlup principle (OMP)~\cite{Doi19}, the Onsager-Machlup functional (OMF) is minimized 
to determine the most probable trajectory.
The OMP has been used to study long-time behavior in soft matter~\cite{Doi19}, non-equilibrium statistical 
physics~\cite{Taniguchi07,Taniguchi08}, structural transitions in protein folding~\cite{ZuckermanBook}, chemical 
kinetic models~\cite{Wang10}, and transitions between laminar and turbulent flow~\cite{Hiruta26}.  
Some of the present authors have used the OMP to study Brownian motion in a fluctuating fluid~\cite{Yasuda24}, 
non-reciprocal stochastic systems~\cite{Yasuda21JPSJ,Yasuda23}, 
active Brownian particles~\cite{Yasuda22,Zheng25},
and informational active matter with observation and feedback mechanisms~\cite{Yasuda25}.

Despite the remarkable success of the OP and OMP, several limitations have been recognized.
For example, the close correspondence of these principles with analytical mechanics has remained unsatisfactory; 
in particular, the covariance structure characteristic of the Lagrangian formulation was not made clear.
This is exemplified by recent questioning of the invariance of the (restricted) free energy under 
a change of variables~\cite{Uneyama20,Nakamura24}.
Furthermore, because the original OP and OMP were formulated for purely dissipative systems, they do not provide a 
variational formulation for the dynamics with inertia.
Although the time derivative of the kinetic energy has conventionally been incorporated into the Rayleighian 
for inertial systems~\cite{Wang2020,Xiao2024,Yasuda25}, 
the covariance of the resulting phenomenological equations has yet to be rigorously examined.

To address the open issues mentioned above, we examine the covariance of the formulations based on the OP and 
OMP throughout the paper and clarify their geometric consistency under coordinate transformations. 
In particular, we consider how the standard OP and OMP can be extended to active and inertial systems, for which such 
formulations have not yet been carefully examined from this perspective. 
In the presence of thermal fluctuations, we also show that the extended OMP is consistent with stochastic thermodynamics~\cite{Sekimoto2010,Peliti2021,Shiraishi2023,Seifert2025}. 
These generalized OP and OMP offer a unified variational framework for deriving the dynamical equations of active and inertial systems.

In Sec.~\ref{thermodynamics}, we define various thermodynamic quantities and introduce a covariant non-equilibrium free energy.
In Sec.~\ref{activeforce}, we discuss the covariant OP and OMP with active forces, and show that the 
extended OMP is consistent with the detailed fluctuation theorem.
In Sec.~\ref{inertia}, we extend the OP and OMP to incorporate inertial effects by introducing the covariant acceleration.
Some examples involving active forces and inertia are presented in Sec.~\ref{examples}.
Section~\ref{summary} summarizes our results and discusses possible applications.

\section{Covariant non-equilibrium free energy}
\label{thermodynamics}

\begin{figure}
\includegraphics[width=0.45 \textwidth,draft=false]{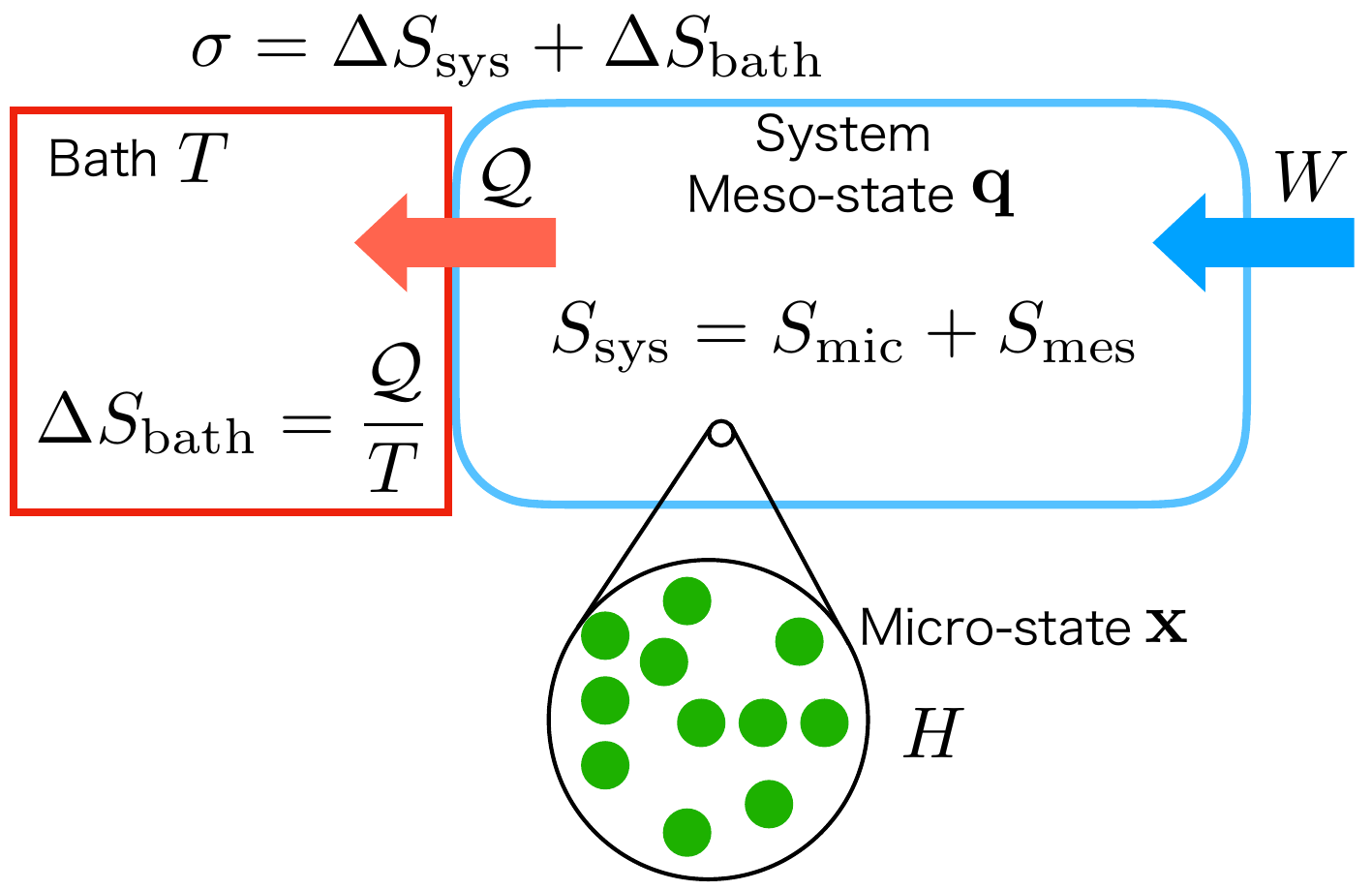}
\caption{
Entropy balance for a system coupled to a heat bath at temperature $T$.
The meso-state of the system is described by $\bfq$, while the microscopic configuration is 
represented by $\bfx$ with Hamiltonian $H$.
The system entropy is decomposed as $S_{\mathrm{sys}}=S_{\mathrm{mic}}+S_{\mathrm{mes}}$.
If the system exchanges heat $\mathcal{Q}$ with the bath and work $W$ with the surroundings, the bath entropy 
change is $\Delta S_{\mathrm{bath}}=\mathcal{Q}/T$, and the total entropy production 
is given by $\sigma=\Delta S_{\mathrm{sys}}+\Delta S_{\mathrm{bath}}$.
}
\label{figure}
\end{figure}

We consider a system in contact with a heat bath at temperature $T$, as shown 
in Fig.~\ref{figure}.
The generalized microscopic coordinates of the system, such as the positions and momenta of all molecules, 
are represented by $\bfx=(x^1,x^2,...)$. 
The state of the system is characterized by a probability distribution function (PDF) $P(\bfx,t)$.
We assume that the system is in a non-equilibrium state, so that $P(\bfx,t)$ is not specified a priori and is generally 
time dependent.
We consider the Hamiltonian $H(\bfx,t)$, which represents the energy of the micro-state $\bfx$ and is also 
time dependent.
In contrast, when the system is in equilibrium, the steady-state PDF is given by the canonical distribution,  
$P(\bfx)\sim {\rm e}^{-H(\bfx)/k_{\rm B}T}$, where $H(\bfx)$ is a time-independent Hamiltonian and $k_{\rm B}$ is the Boltzmann constant
(see Appendix~\ref{App:TD} for details).

In irreversible thermodynamics, we are interested in a coarse-grained level of description and use 
several mesoscopic state (meso-state) variables $\bfq=(q^1,q^2,...)$, rather than in the full micro-state $\bfx $.
We assume that the meso-state variables $\bfq $ are uniquely determined when all of the 
microscopic coordinates $\bfx$ are given, i.e., $\bfq = \bfq(\bfx)$.

Let us consider the local thermodynamic quantities under a given meso-state 
$\bfq$~\cite{esposito2012stochastic}
(see Appendix~\ref{App:GT} for the macroscopic thermodynamics).
First, the system entropy is defined by~\cite{Shiraishi2023,Maruyama09,Parrondo15}
\begin{align}
S_{\rm sys}(\bfq, t)= -k_{\rm B}\int {\rm d}\bfx\, P(\bfx, t|\bfq)\ln P(\bfx, t),
\label{Ssys}
\end{align}
where $P(\bfx, t|\bfq)$ is the conditional PDF of the micro-state $\bfx$ under the 
given meso-state $\bfq$. 
The system entropy $S_\mathrm{sys}$ is a scalar quantity that is 
invariant under the (phase-space preserving) canonical transformation 
of $\bfx$, as justified by Liouville's theorem~\cite{Goldstein2001},  
and under the point transformation of $\bfq$,  i.e., a coordinate transformation in configuration space depending only on position and not on velocity.

The system entropy can be decomposed into microscopic and mesoscopic parts, 
$S_\mathrm {sys}=S_\mathrm{mic}+S_\mathrm{mes}$~\footnote{
Since $\bfq=\bfq(\bfx)$, the microstate probability density can be factorized into a 
conditional density at fixed $\bfq$ and the scalar mesostate density, provided that
compatible reference measures are used. 
Thus, we have $P(\bfx)\propto P(\bfx|\bfq)\rho(\bfq)$. 
Hence, we obtain $\ln P(\bfx) = \ln P(\bfx|\bfq)+\ln\rho(\bfq) +\mathrm{const.}$}, 
where the microscopic entropy is given by
\begin{align}
S_\mathrm{mic}(\bfq, t)=-k_{\rm B}\int \D\bfx\, P(\bfx, t|\bfq)\ln P(\bfx, t|\bfq),
\label{Smic}
\end{align}
which is a scalar with respect to both $\bfx$ and $\bfq$, similar to $S_\mathrm {sys}$.

The mesoscopic entropy, on the other hand, can be written as~\cite{Nakamura24,Graham77,Graham77b,dekker1980path,Yasuda26} 
\begin{align}
S_\mathrm{mes}(\bfq, t)=-k_{\rm B}\ln \rho(\bfq, t)=-k_{\rm B}\ln \left[P(\bfq, t)/\sqrt{g(\bfq)}\right], 
\label{Smac}
\end{align}
where $\rho(\bfq, t)$ is the scalar probability density of the meso-state $\bfq$, and 
$P(\bfq, t)$ is the PDF of the meso-state $\bfq$.
Moreover, $g(\bfq)=\det g_{ij}$ is the determinant of the metric tensor $g_{ij}$ that represents the geometry of the mesoscopic 
configuration space and defines the distance as $(d \bfq)^2=dq^ig_{ij}dq^j$.
Hereafter, we adopt Einstein's convention for repeated indices, 
whereas the superscript and subscript indicate the contravariant and covariant components, respectively.
Because the scalar probability density $\rho(\bfq, t)$ is invariant under point transformations 
of the meso-state $\bfq$, $S_\mathrm{mes}$ is also a scalar for $\bfq$,
although $P(\bfq, t)$ itself is not a scalar~\cite{Nakamura24,Graham77,Graham77b,dekker1980path,Yasuda26}.

Let us further define the internal energy as~\cite{Parrondo15}
\begin{align}
U(\bfq, t)=\int \D \bfx\, P(\bfx, t|\bfq) H(\bfx, t).
\label{internal}
\end{align}
As shown in Fig.~\ref{figure}, the heat dissipated into the heat bath $\mathcal{Q}=\int \delta \mathcal Q$
and the work $W=\int \delta W$ done on the system are given by 
\begin{align}
\delta \mathcal{Q}&=-{\rm d}t\int \D \bfx\, \partial_t P(\bfx, t|\bfq) H(\bfx, t),
\label{Qheat}
\\
\delta W&={\rm d}q^i\int \D \bfx\, \partial_i P(\bfx, t|\bfq) H(\bfx, t)+{\rm d}t\int \D \bfx\, P(\bfx, t|\bfq) \partial_t H(\bfx, t),
\label{Eq:work}
\end{align}
where $\partial_i=\partial /\partial q^i$ and $\partial_t=\partial /\partial t$.
The first law of thermodynamics states that $\Delta U = - \mathcal{Q}+W$~\cite{Sekimoto2010}, 
where $\Delta U$ is the change in the internal energy.
Notice that the entropy change of the bath is given by $\Delta S_{\rm bath}=\mathcal{Q}/T$.

Finally, we define the non-equilibrium free energy of the system as 
\begin{align}
F(\bfq, t)=U(\bfq, t)-TS_\mathrm{mic}(\bfq, t).
\label{Eq:freeEn}
\end{align}
Since we use here the microscopic entropy $S_\mathrm{mic}$ defined in Eq.~(\ref{Smic}),  
$F(\bfq, t)$ is a scalar under point transformations.  
The entropy production $\sigma$, which is the total entropy change in the system and bath,
is given by $\sigma=\Delta S_\mathrm {sys}+\Delta S_\mathrm {bath}
=\Delta S_\mathrm {mic}+\Delta S_\mathrm{mes}+\Delta S_{\mathrm{bath}}$. 
In terms of $F$ and $W$, $\sigma$ can also be rewritten as 
\begin{align}
\sigma = \frac{W}{T} -\frac{\Delta F}{T} +\Delta S_\mathrm{mes},
\label{firstlaw}
\end{align}
where $\Delta F$ is the free energy change, and we have used the first law mentioned above.

\section{Variational principles with active forces}
\label{activeforce}

\subsection{Onsager principle with active forces}

According to the second law of thermodynamics, meso-state variables $\bfq$ 
evolve in such a way that the average entropy production $\sigma$ is non-negative. 
It is known that such irreversible dynamics can be formulated by the 
OP~\cite{Onsager31a,Onsager31b,DoiBook}.
In the OP for active systems, we consider a scalar Rayleighian defined as~\cite{Wang21,Zheng25} 
\begin{align}
R(\bfq, \dot{\bfq})=\Phi(\bfq, \dot{\bfq})+
\dot F(\bfq, \dot{\bfq})-\dot W(\bfq, \dot{\bfq}),
\label{Eq:Ray}
\end{align}
where the dot indicates the time derivative.
Although the Rayleighian can be generalized to an explicitly time-dependent form, $R(\bfq,\dot{\bfq},t)$, for example 
through the time-dependent free energy in Eq.~(\ref{Eq:freeEn}), we restrict ourselves below to the case without explicit time dependence 
in order to focus on the covariance of the OP and OMP.
In Eq.~(\ref{Eq:Ray}), $\Phi$ is the dissipation function given 
by a quadratic form in the generalized velocities $\dot {\bfq}$,
$\Phi(\bfq,\dot{ \bfq})=\dot q^i \zeta_{ij}(\bfq)\dot q^j/2$, where $\zeta_{ij}(\bfq)$ 
is the state-dependent resistance tensor~\cite{DoiBook}.
Therefore, $\Phi$ itself is a scalar. 
The resistance tensor should be symmetric and positive-definite due to Onsager's reciprocal relations
and the second law of thermodynamics~\cite{Onsager31a,Onsager31b}.

Although the definitions of the work $W$ and free energy $F$ are given in Eqs.~(\ref{Eq:work}) 
and (\ref{Eq:freeEn}), respectively, they are generally expressed as functions of the 
meso-state variables $\bfq$ by considering the specific symmetry of the system.
Examples include the Flory-Huggins free energy 
for polymer  solutions~\cite{vanSaarloos2024SoftMatter,DoiBook}, the Frank elasticity of 
nematics~\cite{vanSaarloos2024SoftMatter,DoiBook}, and the Helfrich elastic energy of membranes~\cite{Fournier15,Okamoto16,vanSaarloos2024SoftMatter}. 
Once the scalar free energy $F(\bfq)$ is given as a function of $\bfq$, its 
rate of change is obtained as $\dot F(\bfq,\dot {\bfq})=\dot q^i\partial_i F(\bfq)$.

Furthermore, the power $\dot W$ can be expressed as the product of the state-dependent external or 
active forces $X_i(\bfq)$ and the generalized velocities, i.e., 
$\dot W(\bfq,\dot {\bfq})= \dot q^i X_i(\bfq)$~\cite{Wang21}. 
Note that $X_i$ is a covariant vector and, hence, $\dot W$ is a scalar.
Several studies have pointed out that $\dot{W}$ can describe not only external driving but also internal driving mechanisms, 
as in active Brownian particles~\cite{Zheng25} (see Sec.~\ref{exampleA} below), active nematics~\cite{Wang21,Yasuda24}, active stress~\cite{Wang21}, 
and systems with non-reciprocal interactions~\cite{Yasuda22B,Yasuda21JPSJ}.
This contribution to the Rayleighian is important for applying the OP for active matter. 
In the absence of the active power $\dot{W}$, the Rayleighian of Eq.~(\ref{Eq:Ray}) reduces to 
the one discussed by Doi~\cite{DoiBook}.

With these definitions, the Rayleighian is a scalar under point transformations of $\bfq $. 
The OP states that the instantaneous velocity $\dot {\bfq}$ of the meso-state $\bfq$
is determined by the following phenomenological equation~\cite{DoiBook}
\begin{align}
\left(\frac{\partial R(\bfq, \dot{\bfq})}{\partial \dot q^i}\right)_{\bfq}=0,
\label{Eq:OVP}
\end{align}
where $\bfq$ is held fixed in the partial derivative. 
Substituting the Rayleighian of Eq.~(\ref{Eq:Ray}) into Eq.~(\ref{Eq:OVP}), we obtain the following dynamical 
equation~\footnote{
When the free energy depends explicitly on time, $F(\bfq,t)$, as in Eq.~(\ref{Eq:freeEn}), 
we have $\dot{F}(\bfq, \dot{\bfq},t)=\dot q^i\partial_i F(\bfq, t) + \partial_t F(\bfq,t)$. 
However, since the second term does not depend on $\dot q^i$, it does not contribute to the variational equation. 
Hence, the resulting dynamical equation remains unchanged, except that $F(\bfq)$ is replaced by $F(\bfq,t)$.}
\begin{align}
\zeta_{ij}(\bfq)\dot q^j+\partial_i F(\bfq)-X_i(\bfq)=0.
\end{align}
Notice that we have used here Onsager's reciprocal relations $\zeta_{ij}=\zeta_{ji}$~\cite{Onsager31a,Onsager31b,DoiBook}.

We show that this phenomenological equation is covariant under point transformations. 
Consider a coordinate transformation from $\bfq$ to $\bfQ$
that is related to the original coordinate $\bfq$ by the function 
$\bfQ(\bfq)$. 
We assume that the original coordinate can be represented by the inverse function $\bfq(\bfQ)$. 
Using the chain rule of partial derivatives, one can show the following identity~\cite{Goldstein2001} 
\begin{align}
\dot q^i(\bfQ, \dot{\bfQ})=\frac{\partial q^i}{\partial Q^j}\dot Q^j. 
\label{transformation}
\end{align}
Therefore, in the new coordinates $\bfQ$, the phenomenological equation~(\ref{Eq:OVP})  is also written as 
\begin{align}
\left(\frac{\partial R(\bfQ, \dot{\bfQ})}{\partial \dot Q^i}\right)_{\bfQ}
=\left(\frac{\partial R(\bfq, \dot{\bfq})}{\partial \dot q^j}\right)_{\bfq}
\left(\frac{\partial \dot q^j}{\partial \dot Q^i}\right)_{\bfQ}=0.
\end{align}
Hence, the phenomenological equation is covariant under point transformations.

\subsection{Onsager-Machlup principle with active forces}

To describe the properties of thermal fluctuations around the phenomenological equation, 
we  consider the following Onsager-Machlup Lagrangian (OML)~\cite{Doi19}
\begin{align}
L(\bfq,\dot{\bfq})=R(\bfq,\dot{\bfq})-R_{\rm min}(\bfq),
\label{OML}
\end{align}
where $R$ is the Rayleighian defined in Eq.~(\ref{Eq:Ray}), and $R_{\rm min}$ 
is its minimal value with respect to the generalized velocity $\dot{\bfq}$ 
\begin{align}
R_{\rm min}(\bfq)=\min_{\dot {\bfq}|\bfq} R(\bfq,\dot{\bfq}).
\label{Rmin}
\end{align}
To determine $R_{\rm min}$, $\bfq$ is fixed as in Eq.~(\ref{Eq:OVP}). 
The term $R_{\rm min}$ is necessary in Eq.~(\ref{OML}) to satisfy $L=0$ when the path follows the
phenomenological equation~(\ref{Eq:OVP}). 
The OML in Eq.~(\ref{OML}) is a function of $\bfq$ and $\dot{\bfq}$,
and we define the Onsager-Machlup functional (OMF) as~\cite{Onsager53,Machlup53b,Doi19}  
\begin{align}
O[\bfq(t)]=\frac{1}{2k_{\rm B}T}\int_0^\tau \D t\, L(\bfq(t),\dot{\bfq}(t)),
\label{Eq:OM}
\end{align}
where $\tau$ denotes the final time.

Consider the multi-time joint probability of observing states $\bfq_i$ at times $t_i$ for  
$i=0, 1,2,\dots,n$ under the initial condition $\bfq_0$,
\begin{align}
\rho(\{\bfq_i\}|\bfq_0)=\prod_{i=0}^{n-1}\rho(\bfq_{i+1},t_{i+1}|\bfq_i,t_i),
\end{align}
where $\rho(\bfq_{i+1},t_{i+1}|\bfq_i,t_i)$ is the scalar probability density of 
$\bfq_{i+1}$ at $t_{i+1}$ conditioned on $\bfq_i$ at $t_i$, for the time interval 
$\Delta t= \tau/n$.
Taking the limit of $\Delta t\to0$ and $n \to \infty$ while keeping $\tau$ fixed,
we obtain the conditioned path probability density of the meso-state trajectory $\bfq(t)$ 
under the initial condition $\bfq_0$ as 
\begin{align}
\rho[\bfq(t) |\bfq_0]=\lim_{\Delta t\to 0}\rho(\{\bfq_i\}|\bfq_0),
\end{align}
up to the final time $\tau$.

By using the OMF in Eq.~(\ref{Eq:OM}), the path probability density is given by~\cite{Onsager53,Machlup53b}
\begin{align}
\rho[\bfq(t)|\bfq_0]=N\exp(-O[\bfq(t) ]),
\label{Eq:PathProb}
\end{align}
where $N$ is a normalization factor.
According to OMP, the functional minimization of $O$ in Eq.~(\ref{Eq:OM}) with respect 
to $\bfq(t)$~\cite{Onsager53,Machlup53b}, 
\begin{align}
\frac{\delta O[\bfq(t)]}{\delta q^i}=0,
\end{align}
leads to the following Euler-Lagrange equation for the most probable path 
\begin{align}
\left(\frac{\partial L(\bfq,\dot{\bfq})}{\partial q^i}\right)_{\mathbf {\dot q}}-
\frac{\D} {\D t}\left(\frac{\partial L(\bfq,\dot{\bfq})}{\partial \dot q^i}\right )_{ \bfq}=0.
\label{EulerLagrange}
\end{align}
The covariance of the Euler-Lagrange equation always holds, as is well known in classical 
mechanics~\cite{Goldstein2001}. 
The obtained most probable path can be used, for example, in the saddle-point approximation of the
OMF~\cite{Yasuda24}.

\subsection{Detailed fluctuation theorem}

We now examine the connection between the OMF and stochastic 
thermodynamics~\cite{Shiraishi2023,Seifert2025}. 
According to the detailed fluctuation theorem, the entropy production $\sigma$ is given by~\cite{Seifert12}
\begin{align}
\sigma=k_{\rm B}\ln\frac{\rho[\bfq(t)]}{\rho[\bfq_\mathrm{rev}(t)]},
\label{Eq:FT}
\end{align}
where the system evolves along the path $\bfq(t)$ over the time duration $\tau$. 
Here, $\rho[\bfq(t)]=\rho[\bfq(t)|\bfq_0]\rho(\bfq_0,0)$ is the 
unconditioned path probability density and $\bfq_\mathrm{rev}(t)=\bfq(\tau-t)$ 
denotes the time-reversed trajectory.
From Eq.~(\ref{Eq:FT}), one can derive the integral fluctuation theorem~\cite{Shiraishi2023,Seifert2025} 
\begin{align}
\langle {\rm e}^{-\sigma/k_{\rm B}}\rangle=1,
\end{align}
where $\langle \cdots \rangle$ indicates the average over the path probability density $\rho[\bfq(t)]$.  
The second law, $\langle\sigma\rangle\ge0$, immediately follows from  
Jensen's inequality.

Substituting the path probability of Eq.~(\ref{Eq:PathProb}) into Eq.~(\ref{Eq:FT}), and using  
the Rayleighian of Eq.~(\ref{Eq:Ray}), we obtain
\begin{align}
\sigma&=k_{\rm B}\ln\frac{\rho[\bfq(t)|\bfq_0]}{\rho[\bfq_\mathrm{rev}(t)|\bfq_\tau]}
+k_{\rm B}\ln\frac{\rho(\bfq_0,0)}{\rho(\bfq_\tau,\tau)}\nonumber\\
&=\frac{W}{T}-\frac{\Delta F}{T}+\Delta S_{\mathrm{mes}},
\label{Eq:sigma-OM}
\end{align}
where  
$W=\int_0^\tau {\rm d}t \, \dot{W}(\bfq, \dot{\bfq})$ 
and we have used Eq.~(\ref{Smac}) for $S_{\mathrm{mes}}$.
This expression for $\sigma$ coincides with that of Eq.~(\ref{firstlaw}) and indicates that the entropy production 
in Onsager-Machlup theory is equivalent to that in stochastic thermodynamics and 
is consistent with the first law. 
Notice that we have used the time-reversal symmetry relations 
$\Phi(\bfq)=\Phi(\bfq_\mathrm{rev})$ and 
$R_\mathrm {min}(\bfq)=R_\mathrm{min}(\bfq_\mathrm{rev})$.

\section{Variational principles with inertia}
\label{inertia}

\subsection{Onsager principle with inertia}

Next, we develop a covariant OP that incorporates inertia. 
For this purpose, we introduce the rate of change of the kinetic energy $K$ into the Rayleighian 
in the following way~\cite{Wang2020,Xiao2024,Yasuda25} 
\begin{align}
{\mathcal R}(\bfq,\dot {\bfq}, \bfa)=\Phi(\bfq,\dot {\bfq})
+\dot F(\bfq,\dot {\bfq})-\dot W(\bfq,\dot {\bfq}) +\dot K(\bfq,\dot {\bfq}, \bfa).
\label{Eq:Ray-K}
\end{align}
In terms of $\dot {\bfq}$, the kinetic energy is given by $K=\dot q^i \mu_{ij}(\bfq)\dot q^j/2$, 
where $\mu_{ij}(\bfq)$ is the inertia tensor that is symmetric and positive-definite. 
Hence, $K$ is also a scalar.
By noting $ \mu_{ij}= \mu_{ji}$, 
the time derivative of the kinetic energy becomes 
$\dot K(\bfq,\dot {\bfq}, \bfa)=\dot q^i \mu_{ij}(\bfq) a^j$, 
where the covariant acceleration is given by~\cite{Wald84}
\begin{align}
a^i=\ddot q^i+\Gamma_{jk}^i\dot q^j\dot q^k,
\label{covarinata}
\end{align}
and the Christoffel symbols are defined as~\cite{Wald84}
\begin{align}
\Gamma_{jk}^i=\frac{1}{2}\mu^{il}\left(\partial_j \mu_{kl}+\partial_k \mu_{jl}-\partial_l \mu_{jk}\right).
\label{Eq:Christoffel}
\end{align}
In the above, $\mu^{ij}$ is the inverse matrix of $\mu_{ij}$, i.e., $\mu^{ij}\mu_{jk}=\delta_k^i$
and $\delta_k^i$ is the Kronecker delta, and $\mu_{ij}$ serves as the metric tensor $g_{ij}$. 
In contrast to the previous section, the Rayleighian ${\mathcal R}$
in Eq.~(\ref{Eq:Ray-K}) is a function of the meso-state variables $\bfq$, 
the velocities $\dot{ \bfq}$, and the covariant accelerations $\bfa$.

At a given state $\bfq$ and $\bfa$, the instantaneous velocity 
$\dot {\bfq}$ is determined by the following phenomenological equation that generalizes Eq.~(\ref{Eq:OVP})
\begin{align}
\left(\frac{\partial {\mathcal R(\bfq,\dot {\bfq}, \bfa)}}{\partial \dot{q}^i}\right)_{\bfq, \bfa}=0,
\label{Eq:OVP-K}
\end{align}
where both $\bfq$ and $\bfa$ are held fixed in the derivative. 
Substituting ${\mathcal R}$ of Eq.~(\ref{Eq:Ray-K}) into Eq.~(\ref{Eq:OVP-K}), we obtain 
\begin{align}
\zeta_{ij}(\bfq)\dot q^j+\partial_i F(\bfq)-X_i(\bfq)+\mu_{ij}(\bfq) a^j=0.
\end{align}
In Eq.~(\ref{Eq:OVP-K}), the partial derivative is taken with respect to $\dot{\bfq}$, 
while $\bfq$ and $\bfa$ are kept fixed. 
This mathematical structure has a clear physical basis: $\dot{\bfq}$ governs the irreversible dynamics because it is odd under time reversal, 
whereas $\bfq$ and $\bfa$ are even. 
Furthermore, this extended OP yields the correct equation of motion, including the dissipative force, even when it is formulated in curvilinear
coordinates (see also Sec.~\ref{exampleB}).

Next, we demonstrate that the covariance of the phenomenological equation is preserved under point transformations. 
We consider a point transformation from the original coordinate $\bfq$ to a new 
coordinate $\bfQ$ via a function $\bfQ(\bfq)$, and assume that its inverse $\bfq(\bfQ)$ exists.
While the velocity $\dot {\mathbf q}$ transforms as a contravariant vector via Eq.~(\ref{transformation}), the standard 
acceleration $\ddot q^i$ transforms as 
$\ddot q^i (\bfQ, \dot{\bfQ}, \ddot{\bfQ})
= (\partial q^i/\partial Q^j) \ddot Q^j + [\partial^2 q^i/(\partial Q^j \partial Q^k)] 
\dot Q^j \dot Q^k$, which is neither covariant nor contravariant.
In contrast to the standard acceleration, the covariant acceleration transforms as a contravariant vector: 
\begin{align}
a^i(\bfQ, \bfA) = \frac{\partial q^i}{\partial Q^j} A^j,
\end{align}
where the covariant acceleration in the new coordinates is 
$A^i = \ddot Q^i + \bar\Gamma_{jk}^i \dot Q^j \dot Q^k$, and $\bar\Gamma_{jk}^i$ are the 
Christoffel symbols [defined in Eq.~(\ref{Eq:Christoffel})] for the new coordinate 
$\bfQ.$

By applying the chain rule, we obtain 
\begin{align}
\left(\frac{\partial {\mathcal R(\bfQ,\dot {\bfQ}, \bfA)}}{\partial \dot{Q}^i}\right)_{\bfQ, \bfA} 
= \left(\frac{\partial {\mathcal R(\bfq,\dot {\bfq}, \bfa)}}
{\partial \dot q^j}\right)_{\bfq, \bfa} \left(\frac{\partial \dot q^j}{\partial \dot Q^i}\right)_{\bfQ} 
= 0,
\end{align}
where we have used Eq.~(\ref{Eq:OVP-K}).
This result shows that the phenomenological equation remains covariant under general point transformations, 
provided that the partial derivative is taken with the covariant acceleration held fixed.

\subsection{Onsager-Machlup principle with inertia}

By extending the discussion of the previous section, the OML with inertia can be defined as~\cite{Machlup53b,Doi19,Taniguchi07,Taniguchi08}  
\begin{align}
{\mathcal L}(\bfq, \dot{\bfq}, \bfa) = {\mathcal R}(\bfq, \dot{\bfq}, \bfa) 
- \mathcal{R}_{\rm min}(\bfq, \bfa),
\label{LagrangianInertia}
\end{align}
where ${\mathcal R}_{\rm min}$ is the minimum value of the Rayleighian with respect to $\dot{\bfq}$
\begin{align}
{\mathcal R}_{\rm min}(\bfq, \bfa) = \min_{\dot {\bfq} |\bfq, \bfa}
{\mathcal R}(\bfq, \dot{\bfq}, \bfa).
\end{align}
Note here that $\bfq$ and $\bfa$ are kept fixed, as in Eq.~(\ref{Eq:OVP-K}). 
The OML with inertia, $\mathcal L$, is now a function of $\bfq$, $\dot{\bfq}$, and $\bfa$.

Then, the OMF with inertia is defined as~\cite{Onsager53,Machlup53b,Doi19} 
\begin{align}
{\mathcal O}[\bfq(t) ] = 
\frac{1}{2k_{\rm B}T} \int_0^\tau \D t\, {\mathcal L}(\bfq(t), \dot{\bfq}(t), \bfa(t)).
\label{OMIinertia}
\end{align}
Following the same argument as before, we obtain the path probability density of the meso-state dynamics $\bfq(t)$ 
under the initial conditions $\bfq_0$ and $\dot{\bfq}_0$ as~\cite{Onsager53,Machlup53b}
\begin{align}
\rho[\bfq(t) | \bfq_0, \dot{\bfq}_0] = 
\mathcal N \exp(-{\mathcal O}[\bfq(t) ]),
\label{Eq:PathProb-K}
\end{align}
where $\mathcal{N}$ is a normalization factor.

According to the OMP,  the functional minimization of $\mathcal O$ in Eq.~(\ref{OMIinertia}) with respect 
to $\bfq(t)$~\cite{Onsager53,Machlup53b},
\begin{align}
\frac{\delta \mathcal O[\bfq(t)]}{\delta q^i}=0,
\end{align}
yields the following Euler-Lagrange equation~\cite{Goldstein2001}
\begin{align}
\left(\frac{\partial {\mathcal L(\bfq, \dot{\bfq}, \ddot{\bfq})}}
{\partial q^i}\right)_{\dot{\bfq}, \ddot{\bfq}}
& - \frac{\D}{\D t} \left(\frac{\partial {\mathcal L(\bfq, \dot{\bfq}, \ddot{\bfq}})}
{\partial \dot{q}^i}\right)_{\bfq, \ddot{\bfq}}
\nonumber \\
& + \frac{\D^2}{\D t^2} \left(\frac{\partial {\mathcal L(\bfq, \dot{\bfq}, \ddot{\bfq}})}
{\partial \ddot{q}^i}\right)_{\bfq, \dot{\bfq}} 
= 0.
\label{ELeqinertia}
\end{align}
We note that $\mathcal L(\bfq, \dot{\bfq}, \ddot{\bfq})$ is obtained 
from Eq.~(\ref{LagrangianInertia}) by using Eq.~(\ref{covarinata}) for $\bfa$.
In Eq.~(\ref{ELeqinertia}), the covariance of the Euler-Lagrange equation is preserved even if the covariant acceleration 
$a^i$ is not explicitly introduced as an independent variable.
This is because the variation can be evaluated by taking partial derivatives with respect to $\ddot{q}^i$.

The consistency of the Onsager-Machlup theory with stochastic thermodynamics is maintained even in the 
presence of inertia.
For inertial dynamics, however, the time-reversed path should be understood with the reversal of all variables
that are odd under time reversal, such as velocities or momenta, 
and the first law of thermodynamics becomes $\Delta U +\Delta K= - \mathcal{Q}+W$.

\section{Examples}
\label{examples}

\subsection{Active Brownian particle}
\label{exampleA}

Here, we demonstrate that the Rayleighian in Eq.~(\ref{Eq:Ray}), which includes the active power term $\dot{W}$, correctly 
describes the dynamics of an active Brownian particle.
Consider a three-dimensional active particle whose position and velocity are given by 
$\mathbf{r}(t)$ and $\mathbf{v}=\dot{\mathbf{r}}$, respectively.
Self-propulsion is introduced by a constant-magnitude velocity $\mathbf{V}(t)=V\mathbf{e}(t)$,
 where $V$ is a constant 
propulsion speed and $\mathbf{e}(t) = (\sin\theta \cos\phi, \sin\theta\sin\phi, \cos\theta)$ is the 
orientational unit vector in spherical coordinates, 
parameterized by the polar and azimuthal angles $\theta(t)$ and $\phi(t)$.
The kinematics of $\mathbf{e}$ is given by $\dot{\mathbf{e}}=\bm{\omega}\times\mathbf{e}$, where 
$\bm{\omega}$ is the angular velocity vector~\cite{DoiBook}.

The Rayleighian of the active particle is given by  $R = \Phi - \dot{W}$, where 
$\Phi = \zeta_{\rm t}\mathbf{v}^2/2 + \zeta_{\rm r}\bm{\omega}^2/2$ is the dissipation function characterized by 
the translational and rotational friction coefficients, $\zeta_{\rm t}$, and
$\zeta_{\rm r}$, respectively~\cite{Romanczuk2012}.
The active power due to self-propulsion is $\dot{W} = \zeta_{\rm t} \mathbf{V} \cdot \mathbf{v}$.
Minimization of $R$ with respect 
to $\mathbf{v}$ and $\bm{\omega}$ yields the deterministic kinematics given by $\mathbf{v}=\mathbf{V}$ and 
$\bm{\omega}=\mathbf{0}$. 
The minimum value of the Rayleighian in Eq.~(\ref{Rmin}) is obtained as $R_{\min}= - \zeta_{\rm t} V^2/2$.

In the presence of thermal fluctuations, the active Brownian particle executes a stochastic trajectory. 
By constructing the OML, $L=R-R_\mathrm{min}$, we obtain the OMF in Eq.~(\ref{Eq:OM}) as 
\begin{align}
O[\mathbf{r}(t),\mathbf{e}(t)] 
=\frac{1}{2k_{\rm B}T}  \int_0^{\tau} \D t \, \left[\frac{\zeta_{\rm t}}{2} \left(\mathbf{v}-\mathbf{V}\right)^2 
+ \frac{\zeta_{\rm r}}{2}\bm{\omega}^2\right].
\label{OMI}
\end{align}
The most probable path of an active Brownian particle can be obtained by minimizing this OMF
and solving the corresponding Euler-Lagrange equation of Eq.~(\ref{EulerLagrange}).
Further details are shown in Refs.~\cite{Yasuda22,Zheng25}.

\subsection{Polar coordinates}
\label{exampleB}

As an example illustrating the covariance of the phenomenological equation derived from the OP with inertia, 
Eq.~(\ref{Eq:OVP-K}), we consider the motion of a particle of mass $\mu$ described in two-dimensional polar coordinates.
The position of the particle is denoted by the radial distance $r(t)$ and the polar angle $\theta(t)$.
The dissipation function and the kinetic energy are given by
\begin{align}
\Phi & =\frac{\zeta}{2}\left (g_{rr} \dot r^2+2 g_{r\theta}\dot r\dot \theta+g_{\theta\theta}\dot\theta^2 \right),
\\
K& =\frac{\mu}{2}\left(g_{rr} \dot r^2+2g_{r\theta}\dot r\dot \theta+g_{\theta\theta}\dot\theta^2\right),
\label{kinetic}
\end{align}
where $\zeta$ is the friction coefficient and $g_{ij}$ is the metric tensor with components 
$g_{rr}=1, g_{r\theta}=0,$ and $g_{\theta\theta}=r^2$. 
Moreover, we assume the presence of a scalar free energy, $F(r,\theta)$, and the absence of active or 
external work, $\dot{W} = 0$.

From Eq.~(\ref{kinetic}), we have $\dot K=\mu(\dot r g_{rr}a^r+\dot \theta g_{\theta\theta}a^\theta)$, 
where the covariant accelerations are $a^r=\ddot r-r\dot\theta^2$ and 
$a^\theta=\ddot\theta+2 \dot r\dot \theta/r$. 
By minimizing the Rayleighian $\mathcal{R}=\Phi + \dot{F} + \dot{K}$, the phenomenological equations
of Eq.~(\ref{Eq:OVP-K}) are obtained as 
\begin{align}
\left(\frac{\partial {\mathcal R}}{\partial \dot r}\right)_{a^r,a^\theta} 
& =\zeta\dot r+\partial_r F+\mu a^r=0,\\
\left(\frac{\partial {\mathcal R}}{\partial \dot \theta}\right)_{a^r,a^\theta} & 
=\zeta r^2\dot \theta+\partial_\theta F+\mu r^2a^\theta=0.
\end{align}
These results are consistent with the equations of motion in polar coordinates. 
If we had fixed $\ddot r$ and $\ddot \theta$ instead of $a^r$ and $a^\theta$, we would have arrived at incorrect equations.

\section{Summary and discussion}
\label{summary}

In this paper, we have discussed the covariance of the generalized 
Onsager principle (OP) and the Onsager-Machlup principle (OMP) for active and inertial systems, 
ensuring geometric consistency under coordinate transformations.
Dissipative dynamics are described by the OP, where minimization of the Rayleighian gives the 
phenomenological equation as in Eq.~(\ref{Eq:OVP}). 
Thermal fluctuations are further incorporated through the path probability defined by the 
Onsager-Machlup functional (OMF) in Eq.~(\ref{Eq:OM}). 
By requiring that the path probability obeys the detailed fluctuation theorem, we have confirmed that the 
extended Onsager-Machlup theory is consistent with stochastic thermodynamics.
For the dynamical equations with inertia, we also showed that the covariant acceleration 
must be held fixed when minimizing the Rayleighian, as shown in Eq.~(\ref{Eq:OVP-K}).

Variational principles, including the OP and OMP, should be covariant 
in the sense that their mathematical structure remains unchanged under point transformations. 
This is a major advantage over approaches that directly write down the dynamical equations.
Previously, Uneyama pointed out that the free energy in the Rayleighian is generally not a scalar under 
coordinate transformations~\cite{Uneyama20}.
However, this issue was resolved by Nakamura~\cite{Nakamura24}, who reconsidered the definition of the canonical 
ensemble in terms of the scalar probability density
$\rho(\bfq, t)$. 
Historically, this subtle issue was also noted by Graham and 
others~\cite{Graham77,Graham77b,dekker1980path}.

All covariance statements in this paper refer to smooth one-to-one point transformations $\bfQ = \bfQ(\bfq)$ on a fixed reduced-state manifold. 
Under such a reparametrization, only the coordinate description changes, whereas the reduced probability measure itself remains unchanged:
$\D P = \rho(\bfq)\sqrt{g(\bfq)} \, \D \bfq = \rho(\bfQ)\sqrt{G(\bfQ)} \, \D \bfQ$.
Accordingly, the scalar density $\rho$, and therefore the free energy $F$ defined from scalar local thermodynamic quantities, are invariant. 
In contrast, the coordinate-dependent probability density 
$P(\bfq,t) =  \rho(\bfq,t)\sqrt{g(\bfq)}$ 
is not invariant.

It is useful to distinguish this situation from other variable changes that also arise in Onsager-type variational formulations. 
As discussed in Sec.~II.B of Ref.~\cite{Doi19}, a parametrization of the form $\bfq=\bfq(\bm{\alpha})$ defines collective coordinates on a chosen trial manifold,
where $\bm{\alpha}$ denotes the parameters.
It should therefore be understood as an approximation scheme, rather than as a coordinate transformation of the entire reduced-state space.
Likewise, introducing auxiliary velocities or fluxes in an extended space subject to kinematic constraints, as in Sec.~II.E of Ref.~\cite{Doi19}, 
merely changes the representation of the dissipation function and does not modify the underlying reduced-state manifold itself.
By contrast, a non-bijective mapping between reduced-state variables represents a genuine change in the coarse graining~\cite{Lin23}. 
In that case, because hidden variables have been integrated out, the reduced entropy and the reduced free energy landscape are generally not connected simply by a Jacobian factor.

In this sense, the present formulation complements Uneyama's covariant formulation based on dissipation~\cite{Uneyama20},
Nakamura's invariant free energy landscape~\cite{Nakamura24}, and Doi's approaches based on trial coordinates and constrained fluxes~\cite{Doi19}. 
Its main advantage lies in clearly distinguishing between coordinate covariance, constrained representations, and genuine changes in the level 
of coarse graining.

The present covariant variational formulation suggests several directions for future work. 
An important next step is to extend the framework to field variables and many-body systems, 
where geometric constraints, collective modes, and nontrivial dissipation may play essential roles~\cite{Nardini2017}. 
It would also be interesting to combine the present formulation with information-driven or feedback-controlled 
dynamics~\cite{Yasuda25,VanSaders26}, where activity, inertia, and information flow may be treated within a unified 
covariant framework. 
These extensions would further clarify the role of covariance in variational formulations of non-equilibrium dynamics.

\begin{acknowledgements}

We would like to express our deep appreciation and great respect for 
Prof.\ Masao Doi, with whom some of us had the honor of collaborating. 
His tremendous contributions to soft matter and polymer physics are seminal and have helped shape the entire field.
We thank Profs.\ C.-X.\ Wu and T.\ Uneyama for useful discussions.

K.Y.\ acknowledges JSPS KAKENHI for Grant-in-Aid for Early-Career Scientists (Grant No.\ 25K17357). 
K.I.\ acknowledges the Japan Science and Technology Agency (JST), FOREST (Grant No.\ JPMJFR212N), and CREST (Grant No.\ JPMJCR25Q1).
D.A.\ acknowledges partial support from the Israel
Science Foundation (ISF) under grant No.\ 226/24. 
S.K.\ acknowledges the support from the National Natural Science Foundation of China (Grant No.\ 12274098) and 
from the Zhejiang Key Laboratory of Soft Matter Biomedical Materials (2025ZY01036 and 2025E10072).
This work was also supported by the JSPS Core-to-Core Program 
``Advanced core-to-core network for the physics of self-organizing active matter" (JPJSCCA20230002).
\end{acknowledgements}

\appendix
\section{Local equilibrium}
\label{App:TD}

In non-equilibrium states, the PDFs $P(\bfx,t)$, $P(\bfx,t|\bfq)$, and 
$\rho(\bfq,t)$ are not specified a priori.
In local equilibrium, where the microscopic coordinates $\bfx $ are in equilibrium under a given 
meso-state $\bfq$, the PDF is given by 
\begin{align}
P(\bfx, t)=\frac{{\rm e}^{-H(\bfx)/k_{\rm B} T}}{\Omega(\bfq(\bfx),t)}.
\end{align}
Here, $H(\bfx)$ is the time-independent Hamiltonian and $\Omega(\bfq(\bfx),t)$ satisfies the normalization condition 
$\int \D \bfx\, P(\bfx, t)=1$.
In this situation, we consider the restricted partition function defined as~\cite{DoiBook} 
\begin{align}
Z(\bfq)=\frac{1}{\sqrt{g}} \int \D \bfx\, {\rm e}^{-H(\bfx)/k_{\rm B}T}\delta(\bfq-\bfq(\bfx)).
\label{partition}
\end{align}
Then, the scalar probability distribution of the meso-state is given by 
$\rho(\bfq, t)= Z(\bfq)/\Omega(\bfq, t)$.
The metric factor $1/\sqrt{g}$ in Eq.~(\ref{partition}) ensures that $Z(\bfq)$ is a scalar, in contrast to the previous 
definition that did not include the metric~\cite{Uneyama20,DoiBook}.

Similar to the footnote in Sec.~\ref{thermodynamics}, 
the conditional probability distribution can be  estimated as 
$P(\bfx|\bfq)
\sim P(\bfx, t)/\rho(\bfq, t) 
= {\rm e}^{-H(\bfx)/k_{\rm B} T}/Z(\bfq)$, which is time independent.
In local equilibrium, the microscopic entropy in Eq.~(\ref{Smic}) can be written as 
\begin{align}
S_\mathrm {mic}(\bfq)&=k_{\rm B}\int \D \bfx\, P(\bfx|\bfq)\ln Z(\bfq)
+\int \D \bfx\, P(\bfx|\bfq) \frac{H(\bfx)}{T}
\nonumber\\
&=k_{\rm B}\ln Z(\bfq)+\frac{U(\bfq)}{T}.
\end{align}
From Eq.~(\ref{Eq:freeEn}), the free energy is given by the restricted partition function as
\begin{align}
F(\bfq)=U(\bfq)-TS_\mathrm{mic}(\bfq)=-k_{\rm B}T\ln Z(\bfq).
\end{align}
This corresponds to the restricted free energy in local equilibrium~\cite{DoiBook}.

In global equilibrium, on the other hand, the microscopic PDF obeys the 
canonical distribution, 
$P(\bfx)={\rm e}^{-H(\bfx)/k_{\rm B} T}/Z$, where $Z=\int \D \bfx\, {\rm e}^{-H(\bfx)/k_{\rm B} T}$ 
is the partition function. 
In this situation, $\Omega(\bfq)=Z$ and the mesoscopic probability density can be written in terms of the 
free energy as~\cite{Nakamura24}
\begin{eqnarray}
\rho(\bfq)=Z(\bfq)/Z ={\rm e}^{-F(\bfq)/k_{\rm B}T}/Z.
\end{eqnarray}

\section{Macroscopic thermodynamics}
\label{App:GT}

The macroscopic internal energy and the macroscopic entropy are defined as
\begin{align}
\langle U(t) \rangle &=\int \D \bfq \sqrt{g} \,\rho(\bfq, t)U(\bfq, t),
\\
\langle S_\mathrm {sys}(t) \rangle &=\int \D \bfq \sqrt{g}\, \rho(\bfq, t)S_\mathrm {sys}(\bfq, t).
\label{averageSsys}
\end{align}
The volume element $\D\bfq \sqrt{g}$ is invariant under the point transformations of $\bfq$. 
Inserting Eq.~(\ref{Ssys}) into Eq.~(\ref{averageSsys}), we obtain 
\begin{align}
\langle S_\mathrm {sys}(t) \rangle = - k_{\rm B} \int \D \bfx \, P(\bfx, t) \ln P(\bfx, t),
\end{align}
which is the macroscopic Shannon entropy.

The macroscopic free energy is given by  
\begin{align}
F_{\rm mac}(t) & = \langle U(t) \rangle - T \langle S_\mathrm {sys}(t) \rangle
\nonumber \\
& = \int \D \bfq \sqrt{g} \, \rho (\bfq, t)
[F(\bfq, t) - T S_\mathrm{mes}(\bfq, t)].
\end{align}
The last equation indicates that the non-equilibrium free energy $F(\bfq, t)$ defined in Eq.~(\ref{Eq:freeEn}) 
plays the role of the internal energy density.

\bibliographystyle{apsrev4-1} 

\end{document}